\documentclass[aps,prb,superscriptaddress,reprint]{revtex4-2}
\usepackage{amsmath,amssymb} 
\usepackage{amsmath} 
\usepackage{amsfonts} 
\usepackage{bm} 
\usepackage[colorlinks=true,citecolor=blue,linkcolor=blue,filecolor=blue,urlcolor=blue]{hyperref}
\usepackage[dvipdfmx]{graphicx}
\usepackage{braket} 
\usepackage{physics}
\usepackage{color}
\usepackage{comment}
\usepackage{mathtools}

\begin{document}
\title{Dynamical dimerization and subdiffusive transport of strongly correlated neutral-ionic systems 
coupled to lattices}
\author{Yuta Sakai}
\author{Chisa Hotta}
\affiliation{The University of Tokyo, Komaba, Meguro-ku, Tokyo 153-8902, Japan}
\begin{abstract}
We study the Monte Carlo dynamics of strongly correlated classical particles coupled to lattice degrees of freedom 
that exhibit the neutral-ionic transitions in one dimension, 
relevant to the organic compounds TTF-CA and TTF-BA. 
The particles carrying up and down spins suffer strong Coulomb interactions and 
undergo a neutral-to-ionic transition when the alternating site potentials become small, 
accompanying the charge transfer to the higher energy sites. 
The model is already shown to exhibit the enhancement of conductivity near the neutral-to-ionic crossover, 
which is carried by the neutral-ionic domain walls (NIDWs) that show diffusive character. 
Here, by incorporating local lattice dimerization effects, the motion of NIDWs becomes subdiffusive, 
and the conductivity is enhanced by one order of magnitude. 
When the ionic and dimerized states coexist, the strong fluctuation of dimer configuration supports the 
antiferromagnetic spin correlation, suppresses Pauli blocking, and subsequently enhances the motion of NIDWs. 
Our results highlight the crucial role of local and fluctuating dimerization in governing 
transport and ferroelectric properties near neutral-ionic transitions.  
\end{abstract}
\maketitle
\section{Introduction}\label{sec:introduction}  
Neutral-ionic(NI) transition is one of the fundamental and old topics in condensed matter physics, 
discussed in analog to the competition between band insulator and Mott insulator. 
In a lattice system consisting of two species of sites or orbitals, the charge transfer from one to the other 
is driven by the subtle balance between electronic interactions and difference in the energy levels. 
When the lattice degrees of freedom come into play, the ionic Mott insulating state loses its inversion symmetry and 
forms a dimer that yields bulk ferroelectricity. 
\par
Organic charge-transfer complex TTF-CA served for a long time as a good platform to study NI transitions 
and related phases \cite{Torrance1981}, where the transitions are controlled by temperature and pressure. 
The collective charge transfer from the donor (D) to an acceptor (A) molecules 
transforms the material from a neutral phase (D$^0$A$^0$) to an ionic phase (D$^+$A$^-$) at around room temperature. 
At lower temperatures, both the neutral and ionic phases show transition into the 
dimerized phase with robust lattice distortion, that breaks the inversion symmetry and drives the ferroelectricity. 
The material is the most popular NI system studied since 1980's 
\cite{Torrance1981,Torrance1981-2,Tokura1982,Tokura1984,Mitani1984,Nagaosa1986-1,Nagaosa1986-2,Nagaosa1986-3,Nagaosa1986-4}, 
and the overall picture mentioned above is known already at around 2000 \cite{Takaoka1987,Cailleau1997}. 
However, it is only recently that the nature of crossover from the neutral to the paraelectric ionic phase 
is experimentally clarified \cite{Takehara2015,Takehara2018}. 
Particularly, there is a very sharp enhancement of the electrical conductivity at the crossover \cite{Takehara2018}, 
which was explained by our previous theory using the same framework as the present paper \cite{Sakai2024}. 
\par
More importantly, this crossover is not merely electronic; it is associated with the reorganization of 
the charge and spin degrees of freedom, 
while it also couples to the local lattice deformations and dipole formation. 
Experiments indicate that the ionic phase at high-pressure-high-temperature is dimerized up to $94 \text{-} 97 \%$ 
\cite{Sunami2018,Sunami2021,Takehara2023}, which is, however, not static, but fluctuates at 
a relatively slow time scale $\sim 10^{11} \rm Hz$, presumably coupled to the motion of charge and spin solitons \cite{Takehara2018}. 
The nuclear quadrupole resonance (NQR) that has the timescale $\sim 10^7 \rm Hz$ slower than this dynamics 
indicate that the system is nondimerized and paraelectric on average. 
Such slow dynamics is very rare \cite{Katayama2021}, and has not been observed in other NI transition materials, 
including $\rm DMTTF \text{-} QCl_4$ \cite{Sunami2019} and $\rm TTF \text{-} QBrCl_3$ \cite{Kagawa2010-2}. 
\par
The types of dimerization observed here do not seem to form a spatial coherence, 
but rather fluctuates independently, possibly because the system is at a high enough temperature 
where the electrons show incoherent and diffusive transport, which then couples to the lattice locally. 
The standard spin-Peierls picture \cite{Northby1982,Hase1993} breaks down, 
nor its local fluctuation cannot be captured simply by the so-called ``dimer liquid" 
in terms of the gas-liquid-solid transitions about the ionic domains 
analogous to water \cite{Lemee-Cailleau1997,Luty2002,Kishine2004}, studied based on the spin-1 Ising model \cite{Blume1971}. 
\par
In the ionic extended Hubbard model (IEHM) at half-filling, which is the basic model for NI transition, 
theories indicate a strong tendency to form a spontaneous ``dimerization" even without the lattice effects 
at the very narrow region between the band and Mott insulating phases in the ground state phase diagram 
\cite{Fabrizio1999,Manmana2004,Kampf2003,Wilkens2001,Torio2001,Legeza2006,Tincani2009}. 
Although the charge gap does not close at these transitions, 
the charge susceptibility weakly diverge \cite{Manmana2004,Aligia2004}, 
which shall be related to the enhanced fluctuation of dimerization that penetrates inside 
the Mott and band insulating phases \cite{Yonemitsu2002-2}.
\par
Previously, we studied the strong-coupling limit of the IEHM using Monte Carlo simulations 
to extract the Glauber dynamics of strongly correlated charges near the NI transition \cite{Sakai2024}. 
We identified the diffusive motion of neutral-ionic domain wall (NIDW) excitations, 
which carry a fractional charge and enhance the electrical conductivity near the NI transition point 
due to their increasing number density. 
However, since lattice degrees of freedom are expected to play a crucial role in both the emergence 
of ferroelectricity and the transport properties in this regime, 
it is natural to expect that low-energy excitations and transport are 
altered when the coupling to lattice distortions is taken into account. 
In this paper, we address this issue by incorporating stochastic and local lattice dimerization into the model, 
and clarify how the complex interplay between electronic correlations, lattice distortions, 
and thermal fluctuations shape the resulting phases and their transport properties.
\par
The paper is structured as follows. 
In \S.\ref{sec:model_methods} we explain our model and the formulation 
to extract the transport properties from our Monte Carlo simulations. 
In \S.\ref{sec:results}, 
we show the phase diagram, how the dimerization takes place, and the related subdiffusive dynamics of NIDWs, 
as well as the analysis including the inter-chain coupling that induces the ferroelectric transition. 
The detailed discussions on how to interpret the experimental findings based on our results are expanded in  
\S.\ref{sec:discussion}. 

\begin{figure*}[tbp]
\includegraphics[width=18cm]{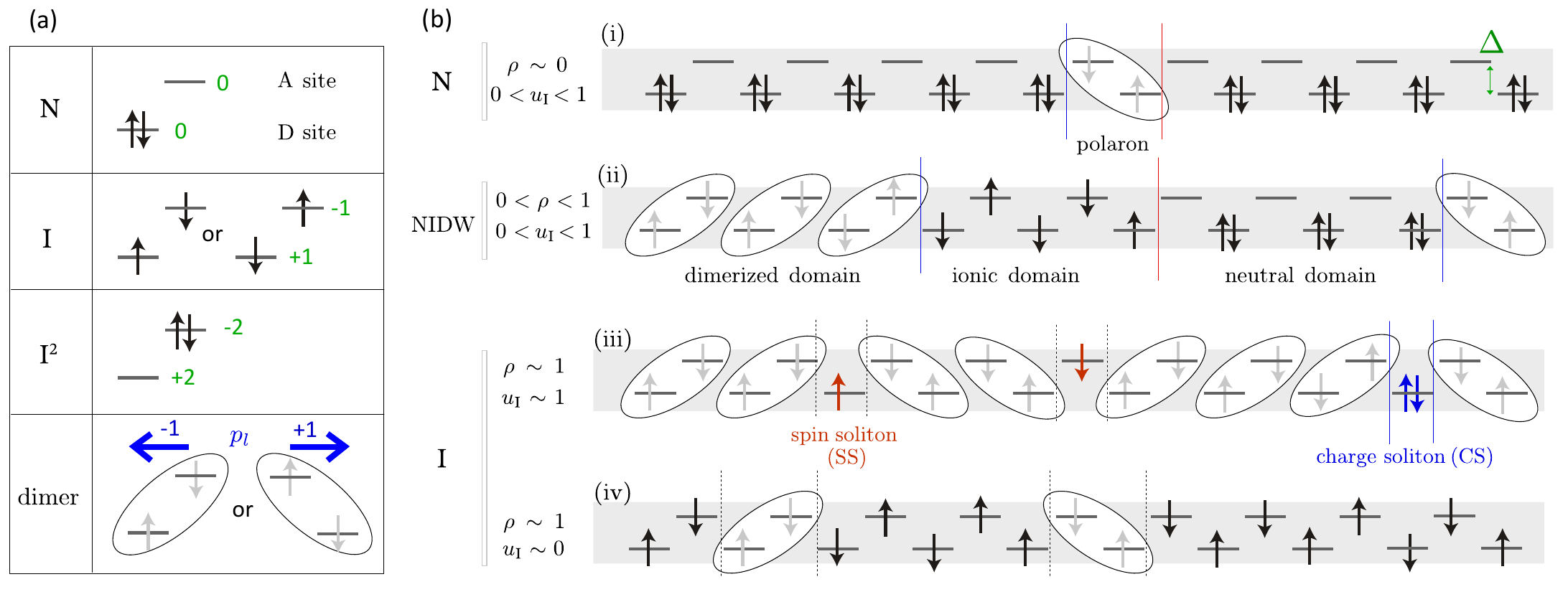}
\caption{
Schematic illustrations of states realized in the NI model. 
(a) Particle configurations, N(neutral), I(ionic), I$^2$(highly ionic) and dimerized ones 
on D and A sites with charge densities measured from the neutral (N) state. 
The dimers carry dipoles pointing toward left $p_l=-1$(DA) or right $p_l=+1$(AD). 
The actual polarization values and directions depend on to what extent it is 
regarded as electronic or displacement types of ferroelectricity. 
(b) Representative states characterized by $\rho$ and $u_{\rm I}$. 
Domain boundaries carrying $+1/2$ and $-1/2$ are shown in red and blue vertical lines, respectively. 
(i) When $\rho\sim 0$ the N state with dimerized neutral pairs of sites may appear as a thermal fluctuation. 
(ii)When $0<\rho<1$ and $0<u_{\rm I}< 1$, dimerized, ionic, and neutral domains appear, 
each spanned over more than two sites. 
(iii) When $\rho \sim 1$ and $u_{\rm I}\sim 1$, most of the ionized states belong to the dimers, 
and the isolated spin sandwiched by dimers behaves as spin soliton (SS), 
and so as the isolated neutral site as charge soliton (CS). 
(iv) When $\rho\sim 1$ and $u_{\rm I} \sim 1$, we have an I state, 
which has a small but finite probability of dimerization that appears as thermal fluctuation. 
  }
\label{fig:model}
\end{figure*}
\section{Model and methods}
\label{sec:model_methods}
\subsection{Model}
\label{subsec:model}
We consider a classical model featuring an electronic system, 
with two species of particles carrying either up or down spins, 
where the two particles with the same spin orientation are not allowed to occupy the same site. 
The most basic form of the model for NI system \cite{Sakai2024} is given as
\begin{align}\label{eq:previous_Hamiltonian}
  \mathcal{H}_{\rm cl}= U\sum_{i=1}^{N}n_{i,\uparrow}n_{i,\downarrow} + V\sum_{i=1}^{N}n_{i}n_{i+1} 
+\Delta \sum_{j\in \text{A}} n_{j}, 
\end{align}
where $N$ is the system size, $n_{i\sigma}=0$ or $1$ is the number of electrons with $\sigma=\uparrow,\downarrow$ spins on the site of $i$ interacting via on-site and nearest neighbor Coulomb interactions, $U$ and $V$, 
and the two sublattices are denoted as A and D, representing the acceptor and donor molecule, respectively. 
We set the energy level of site A to be higher than that of site D by $\Delta$. 
The minimal model for the NI transition had been the 1D ionic extended Hubbard model (IEHM) 
\cite{Nagaosa1986-2} 
with the hopping term, $\sum_{j} -t_r (c_j^\dagger c_{j+1}+\text{h.c.})$, being added to Eq.~(\ref{eq:previous_Hamiltonian}). 
The strong coupling limit of the 1D-IEHM, $t_r/U \rightarrow 0$, naturally yields Eq.~(\ref{eq:previous_Hamiltonian}), 
where the fermionic sign of the IEHM do not matter in taking this limit as far as we consider 
the even number of up/down spin particles. 
In Ref. [\onlinecite{Sakai2024}], the authors investigated this model 
and demonstrated that the transport properties of the NI system in the temperature range of approximately 
$T \sim 0.4-0.5 t_r$ are diffusive, and are well described by this model, 
with a detailed discussion of the validity of such a classical treatment provided therein \cite{Sakai2024}.
\par
The inclusion of the nearest neighbor Coulomb terms is indispensable for the NI transition mechanism in real materials, 
because it originates from the Madelung energy as discussed in Ref. [\onlinecite{Nagaosa1986-2}].
In the present paper, we consider another ingredient, the effect of lattice dimerization, in the model 
to examine the possible realization of the ferroelectric phase and the high-temperature phase with local and dynamical dimerization 
that are observed experimentally. 
For that purpose, we set the Hamiltonian as ${\cal H}={\cal H}_{\rm cl}+{\cal H}_{\rm d}$ with 
\begin{align}
{\cal H}_{\rm d}&= \sum_{i=1}^N {\cal P}_i \big(
-E_d b_{i,i+1} + V_d b_{i-1,i} b_{i,i+1} \notag\\
&+ J (\sigma_i \sigma_{i+1})(1-b_{i-1,i})(1-b_{i,i+1})(1-b_{i+1,i+2}) \big) {\cal P}_{i+1}
\label{eq:hamd}
\end{align}
where $\sigma_i=\pm 1$ denotes the up/down spin degrees of freedom featuring electron spins 
and $\theta$ is the step function. 
We set $b_{i,i+1}=1,0$ as a bond parameter that denotes 
whether the two adjacent sites, $i$ and $i+1$, are dimerized (1) or not (0). 
The projection onto the singly occupied state on the $i$ th site is implied using ${\cal P}_i$. 
The first term represents the net energy gain when site $i$ and $i+1$ are dimerized, 
which phenomenologically represents 
the kinetic energy gain that includes the singlet formation energy and the elastic energy loss. 
When the two adjacent sites are not dimerized, there arises an antiferromagnetic interaction $J$ between them 
which we approximate as Ising term. 
To have a dimerized ground state in the ionic phase, $E_d$ must be positive. 
However, $E_d$ does not necessarily need to take a positive value when the elastic energy loss becomes large, 
which indeed is expected for the TTF-CA in the higher pressure range where the lattice spacings become smaller. 
Even in such case, there can be a total gain of free energy for the states that have a finite number of dimerized bonds 
due to entropic gain. 
We also need to mention that the two adjacent bonds cannot be simultaneously dimerized, 
namely we take $V_d \rightarrow \infty$ throughout this paper. 
\par
For simplicity, we assume that the electric energies, $U, V, \Delta$, are not affected by dimerization. 
Unless otherwise specified, we use the values of $U=7.5$ and $V=3.5$, normalized such that $t_r=0.2$\,eV is a unity  
based on the previous works on TTF-CA \cite{Nagaosa1986-2}. 
We also fix the value, $J={{t_r}^2}/\{2(U-V)\}=0.125$, in the form of second order perturbation 
for the strong coupling case, $t_r, \Delta \ll U-V$. 
While the $E_d$ dependence is examined shortly, it is typically set to $E_d=0.5$, which is comparable to the singlet energy gain, $-3J$. 
The total particle number is set to $N_e=N$, consisting of the same number of particles with up and down spins, 
and apply a periodic boundary condition. 
In our calculation, the system size is taken as $N=256$ at the temperature of $T\sim 0.4 \text{-} 0.6$. 
The size dependence of the simulation is examined in Ref. [\onlinecite{Sakai2024}], which is shown not to change 
the quality of the results. 
In principle, activation energy obtained by the Arrhenius plot of an excitation energy 
does not depend on the temperature in this temperature range \cite{Takehara2019,Takehara2023}. 
%
\subsection{Observables in the thermal equilibrium}
\label{subsec:observables}
Figure~\ref{fig:model}(a) shows schematically the definition of N and I (I$^2$), 
where we count the charge from the neutral state; 
The site has no net charge when it is doubly occupied/empty for D/A site. 
The NIDW denotes the domain that carries $\pm 1/2$ charge, which is defined 
according to the net charge transfer $\pm 1$  between the adjacent two sites measured 
from the neutral state. 
In Fig.~\ref{fig:model}(b), we show the actual NIDWs in vertical solid lines. 
Notice that if the charge transfer is $\pm 2$ the two NIDWs are inserted (not shown).  
\par
The NI transition is characterized by the degree of charge transfer from site D to site A given as, 
\begin{equation}
\rho= 2\sum_{j\in \text{A}} n_{j}/N, 
\end{equation}
which is called ionicity. 
While $\rho$ can range between 0 and 2 in principle, $\rho>1$ is unlikely, 
because the perfect N and I phases have $\rho=0$ and 1, respectively. 
The degree of dimerization in our classical model is defined as 
\begin{equation}
u= 2\sum_{j=1}^{N} b_{j,j+1}/N,
\end{equation}
corresponding to the ratio of dimerized bonds, ranging from $0$ to $1$. 
Here, because $b_{j,j+1}=1$ is realized only when $n_j=n_{j+1}=1$, i.e. the adjacent two sites are ionized, 
the above two quantities are related. 
If $\rho \sim 0$, most of the sites are neutral, 
while there can be a small number of sites that join the dimerization that have $u >0$. 
Even when $n_j=1$ is dominant, whether the dimerization dominates or not changes the nature of the phase. 
\par
As ${\cal H}_{\rm d}$ can thus give variants in the nature of NIDWs, 
we introduce the degree of dimerization normalized by the size of the ionic domain, 
\begin{equation}
\label{eq:ui}
u_{\rm I}=\frac{u}{\rho}. 
\end{equation} 
Figure~\ref{fig:model}(b) shows schematically the representative states realized in the actual calculation. 
If (i) $\rho\sim 0$ and $0<u_{\rm I}< 1$, it means that the system is in a neutral phase, 
including dilute number of dimerization. 
This dimerization may represent the polarons discussed earlier \cite{Fukuyama2016,Tsuchiizu2016}.
When (ii) $0<\rho<1$ and $0<u_{\rm I}<1$, there appear a neutral domain, a dimerized domain, 
and a nondimerized ionic domain extending over more than two sites, that fluctuate in time. 
When (iii) $\rho\sim 1$ and $u_{\rm I} \sim 1$, the dimerized state dominates, in which case 
there appears an isolated ionic site sandwiched by dimers which is called spin soliton (SS), 
whose density is given by $1-u_{\rm I}$. There also appears a charge soliton (CS) 
as isolated neutral site \cite{Fukuyama2016,Tsuchiizu2016}. Both SS and CS are created in pairs. 
Finally, when (iv) $\rho\sim 1$ and $u_{\rm I} \sim 0$, 
most of the sites are ionized, and the thermal fluctuation introduces 
a finite number of isolated dimerized sites, which however, does not develop into a stable domain. 
In the semiclassical approach based on bosonization, 
the quasi-particle excitation in the N phase is the polaron, 
and the ones in the ferroelectric dimerized phase are the CS and SS, 
which are exclusive with each other \cite{Tsuchiizu2016}. 
Case (ii) with complex domain walls is only available here. 
\par
We emphasize the difference between $u$ or $u_{\rm I}$ with 
the degree of dimerization in the previous QMC calculations \cite{Nagaosa1986-3,Otsuka2012}. 
In their model, the dimerization is treated as a spatially homogeneous parameter 
in the form of the modulation of $t_r$, which is optimized at each step of calculation 
by the iterative calculation to minimize the free energy. This kind of treatment features 
a spin-Peierls state, expected at low temperatures where the quantum fluctuations dominate. 
In our case, the dimerization is treated as local and can be induced independently from other parts 
by the thermal fluctuation. 
The parameters $E_d$ and $J$ control the probability of such fluctuation to happen. 
Resultantly, the dimerization can occur both coherently throughout the system or inhomogeneously 
and temporarily, which is expected to happen in the paraelectric ionic phase of TTF-CA and TTF-BA.

\subsection{Monte Carlo algorism}
\label{subsec:algorism}
In the MC algorism, we apply a heat-bath method with the transition probability, 
$P_{\rm bath}(\Delta E)=(1+e^{\Delta E/k_{B}T})^{-1}$, 
where $\Delta E$ is the energy difference between the states before and after 
the trial transition process. 
We perform the simulation by repeating the following MC step (MCS), 
which conserves the number of up and down spin particles. 
A single MCS consists of (1) the electron hopping process 
which is repeated over $N_e/2=N/2$ times per $\sigma=\uparrow,\downarrow$, 
and (2) the introduction/release of dimerization over $N$ times: 
\newline
(1) Randomly choose an electron with spin $\sigma$ and see whether it joins the dimerization. 
\newline
(1-a) If yes, release the dimerization and hop the electron from site A to site D 
with the probability of $P_{\rm bath}$. 
\newline
(1-b) If not, randomly choose a hopping direction and hop the electron with the probability of $P_{\rm bath}$ 
if the destination is empty (for spin-$\sigma$ electron) and the destination site is not dimerized. 
Otherwise, discard the process. 
\newline 
(2) Randomly choose a bond and see whether the bond is dimerized. \newline
(2-a) If dimerized, release the dimerization and assign the spins to the undimerized site 
to either up-down or down-up to have the lowest energy, with the probability of $P_{\rm bath}$. 
If the energy does not depend on the assignment of spins, randomly choose from the two. 
\newline
(2-b) If not dimerized, dimerize the bond with the probability of $P_{\rm bath}$ 
if the neighboring two sites of the bond are each singly occupied with antiparallel spin 
and not joining the dimerization with the other sites. 
Otherwise, discard the process. 
\par
This algorithm is interpreted as the time(MCS) evolution, 
which is so-called Glauber dynamics \cite{Glauber1963}. 
It is generally valid for a system that has no system-encoded intrinsic rule for 
the time evolution of the concerned degrees of freedom. 
In such case, the system is modeled to evolve with stochastic dynamics implicitly induced by 
a weak coupling to a heat bath 
\cite{Binder1992,Binder1997,Binder2022}. 
The examples of the Glauber-type MC dynamics are described in our previous paper 
and the references are therein \cite{Sakai2024}. 

\begin{figure}[htbp]
\includegraphics[width=8.5cm]{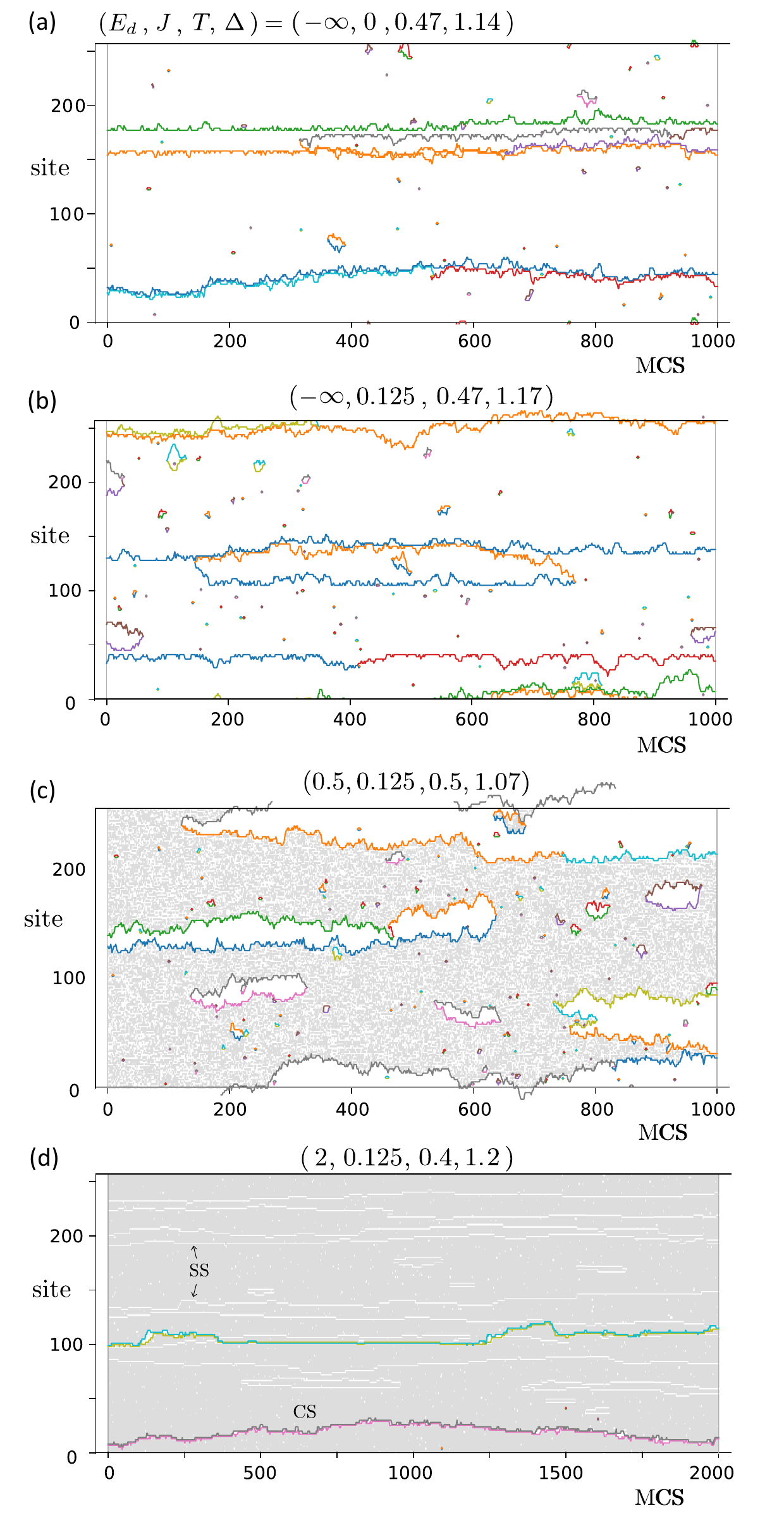}
\caption{
Trajectories of the NIDWs (in colored lines) for four cases, 
(a,b) without dimerization ($E_d=-\infty$) and $J=0,\;0.125$, 
corresponding to (ii) with $u_{\rm I}=0$ in Fig.~\ref{fig:model}, 
(c) $E_d=0.5$ and $J=\;0.125$, corresponding to (ii)  
and (d) $E_d=2$ and $J=0.125$ which is case (iii).  
Gray regions indicate $b_{j,j+1}=1$ (dimerized). 
  }
\label{fig:trajectory}
\end{figure}
\subsection{Analysis}
\label{sec:analysis}
We leverage the dynamics for the two independent analyses on ${\cal H}={\cal H}_{\rm cl}+{\cal H}_{\rm d}$
using the same treatment 
as we did previously for ${\cal H}={\cal H}_{\rm cl}$ \cite{Sakai2024}, and compare these results. 
One is to calculate the electrical conductivity with an electric field bias, 
and the other is to obtain an equilibrium with zero bias,  
calculate the quantities related to diffusions of NIDW, and obtain the conductivity of NIDW that 
carries $\pm 1/2$ charge. 
Here, we briefly summarize the treatment, whose details are discussed more precisely in Ref. [\onlinecite{Sakai2024}]. 
\par
{\it Electric field bias. } 
The external electric field ${\cal E}$ is included in the Hamiltonian as a gradient of potential in the form, 
\begin{equation}
{\cal H}_{{\cal E}}= \sum_{j=1}^N n_j {\cal E} r_j- \sum_{l\in\text{dimer}} \chi p_l {\cal E}, \quad r_j=j. 
\end{equation}
For the dimerized DA and AD bond, assign the polarization $p_l=-1$ and $+1$, respectively. 
In the MC simulation, the bias is reflected in the transition probability 
as such that the particles hopping to the right adds $+{\cal E}$ to $\Delta E$, 
and $-{\cal E}$ for the particles hopping to the left \cite{Kawasaki1966}. 
For the formation/release of dimers, one needs add/reduce $\chi p_l {\cal E}$. 
We performed several calculations and found that small $\chi$ does not influence the result much, 
as they are included in $E_d$ on an average, thus set $\chi=0$ throughout this paper. 
The current ${\cal I}$ is defined as the number of particles passing through 
the periodic boundary divided by the total MCSs of the duration, 
and the conductivity is given as $\sigma_e={\cal I}/{\cal E}$. 
\par
{\it Diffusive NIDW transport. } 
During the MCS evolution, the NIDWs are created in pairs and disappear in pairs when the $+1/2$ and $-1/2$ NIDWs meet. 
After the thermalization process, 
one can track the trajectories of all the NIDWs created/disappeared within a certain time period, $N_{\rm MCS} \sim 10^3$ MCS. 
Figure~\ref{fig:trajectory} shows four cases of the actual trajectories of the NIDWs we calculated for 
${\cal H}={\cal H}_{\rm cl} + {\cal H}_{\rm d}$; 
For the two panels (a) and (b) we set $E_d=-\infty$ to exclude dimerization, 
and include $J=0.125$ for the latter. 
Both corresponds to $u_{\rm I}=0$ case of (ii) in Fig.~\ref{fig:model}(b). 
One finds that in the absence of $J$ the NIDWs once created, 
remain localized near their point of origin. 
In contrast, the presence of $J$ facilitates their spatial propagation. 
When the dimerization effect is included by setting $E_d\gg-\infty$, 
regions with $b_{j,j+1}=1$ begin to emerge, forming broad domains in panel (c), 
that features state (ii). There are finite fluctuations of dimerization inside the dimerized (gray) domain 
that appear in fine white structure. 
In panel (d), the dimerization dominates, corresponding to state (iii), 
and there appear white trajectories that indicate the SS 
as well as the colored CS consisting of a pair of domain walls carrying the same sign of charge. 
\par
Some of the NIDWs recombine after a short lifetime with the same counterpart at 
its creation which forms a small loop, which we call `pair-recombination", 
which are observed as small loops in Fig.~\ref{fig:trajectory}. 
Otherwise, the NIDWs move back and forth. The mean square displacement (MSD) of these NIDWs at time $t$ after 
their creation is evaluated as $\langle x_{t}^{2} \rangle$, and the diffusion constant is given as 
\begin{equation}\label{eq:D} 
  D = \lim_{t\rightarrow\infty} \frac{\langle x_{t}^{2} \rangle}{2t}. 
\end{equation}
\par
The lifetime of NIDWs at time $t$ ($t=0$ is set for each NIDW to the time they are created) 
follows very precisely the distribution function given as,   
\begin{equation}
\label{eq:ndw-fit}
N_{\rm dw}(t) \propto ({\rm const.}+t^{-\frac{3}{2}})e^{-\frac{t}{\tau_{\rm dw}}}. 
\end{equation}
The second term comes from the pair-recombination process, whose number follows $\propto \tau^{-3/2}$, 
known in the 1D random-walk problem. 
By subtracting the second term and focusing only on the contribution from the NIDWs that avoid pair-recombination, 
the number of NIDWs that live longer than $\gtrsim t$ is given by 
\begin{equation}
\sum_{t'=t+1}^\infty N_{\rm dw}(t) \rightarrow {N}_{\rm dw}^{\rm tot} e^{-\frac{t}{\tau_{\rm dw}}}, \quad t\gtrsim \tau_{\rm dw}. 
\end{equation}
where $\tau_{\rm ave}$ is the average lifetime, fitted together with ${N}_{\rm dw}^{\rm tot}$. 
Using ${N}_{\rm dw}^{\rm tot}$, one can evaluate the number density of NIDWs as 
\begin{equation}\label{eq:number_density}
  n_{\rm dw} = {N}_{\rm dw}^{\rm tot}\frac{\tau_{\rm ave}}{N_{\rm MCS}} \frac{1}{N}. 
\end{equation}
\par
Using $D$ and $n_{\rm dw}$, the electrical conductivity $\sigma_{\rm dw}$ is calculated 
based on Einstein relation as 
\begin{equation}\label{eq:sigmadw}
 \sigma_{\rm dw} = \frac{n_{\rm dw} D}{4T}. 
\end{equation}
Previously, we studied ${\cal H}_{\rm cl}$ and showed that the diffusive nature of NIDWs was well preserved, 
as $\langle x_{t}^{2} \rangle = 2D t$ is reproduced at $t\gtrsim 10^3$. 
There, the two evaluations, $\sigma_e$ and $\sigma_{\rm dw}$, showed quantitatively 
the same enhancement at the phase boundaries where the N and I phases compete \cite{Sakai2024}. 
In the present calculation, the dimerization effect renders the dynamics of NIDWs to 
follow the subdiffusive nature at longer time scales, in which case, Eq.~(\ref{eq:sigmadw}) does not 
straightforwardly apply.

\section{Results}\label{sec:results}
\subsection{Thermal equilibrium}\label{subsec:thermal_equilibrium}

\begin{figure}[tbp]
\includegraphics[width=8cm]{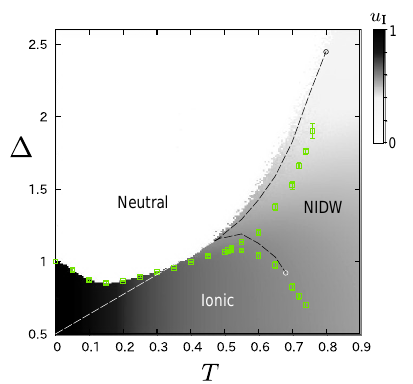}
\caption{
Phase diagram on the $T$-$\Delta$ plane in the thermal equilibrium for ${\cal H}={\cal H}_{\rm cl}+ {\cal H}_{\rm d}$ 
with $U=7.5$, $V=3.5$, $J=0.125$, and $E_d=0.5$, obtained for $N=256$. 
The broken line is the corresponding phase boundary for ${\cal H}_{\rm cl}$ without dimerization. 
The density plot of $u_{\rm I}$ is shown together. 
  }
\label{fig:phase_diagram}
\end{figure}
Figure~\ref{fig:phase_diagram} shows the finite temperature phase diagram on the plane of $\Delta$ and $T$. 
Since $\Delta$ is expected to be a monotonously decreasing function of pressure \cite{Nagaosa1986-2}, 
this phase diagram is approximately regarded as the experimental pressure-temperature one. 
The free energy landscape calculated from the MC histogram has typically a three-minima structure 
in terms of the ionicity $\rho$, 
where the center minimum $0<\rho<1$ hosts a finite concentration of NIDWs as a thermodynamically stable state. 
Using the free energy, we obtained the phase boundaries \cite{Sakai2024}.  
For comparison, we plot together the phase boundaries without dimerization and other parameters 
being the same in the broken line. 
The two diagrams have similar profiles with a triple point, 
and the NIDW phase lies between the homogeneous neutral and ionic phases. 
\par
At higher temperatures, the phase boundary shifts to lower $\Delta$ by $\sim 0.1-0.2$ 
by the introduction of dimerization term, 
which shall be because the neutral phase is relatively stable than the ionic phase against 
the introduction of ${\cal H}_{\rm d}$. 
This shall be because the spins in the ionic phase were previously free spins 
which contributed to the entropy, 
but here, these spins either interact antiferromagnetically or form dimers, 
both working as the reduction of entropy. 
At lower temperatures below the triple point, the slope of the phase boundary differs between the two cases: 
Without dimerization, the straight line is analytically obtained 
from the entropic gain of the ionic phase due to the spin degree of freedom. 
The introduction of dimerization term changes its slope, 
and there arise reentrant types of upturn at low temperature 
and the boundary extrapolates to the zero temperature transition point, $\Delta_c=U-2V+E_d=1$. 
The negative slope of the boundary at $T\lesssim 0.2$ implies that the the entropy of 
the ionic phase is smaller than that of the neutral phase at low temperatures. 
There, the dimerization $b_{j,j+1}=1$ dominates as $u_{\rm I} \sim 1$ 
and the system mimics the ferroelectric phase. 
Generation of the dilute ionic domains (polarons) in the N phase might contribute to the entropic gain. 
The value of $u_{\rm I}$ increases gradually upon lowering $T$ 
inside the ionic phase, and the system crossovers to the ferroelectric ground state. 
The inclusion of inter-chain coupling allows the ferroelectric transition at finite temperature 
as we see shortly. 
\par
\begin{figure}[tbp]
\includegraphics[width=8cm]{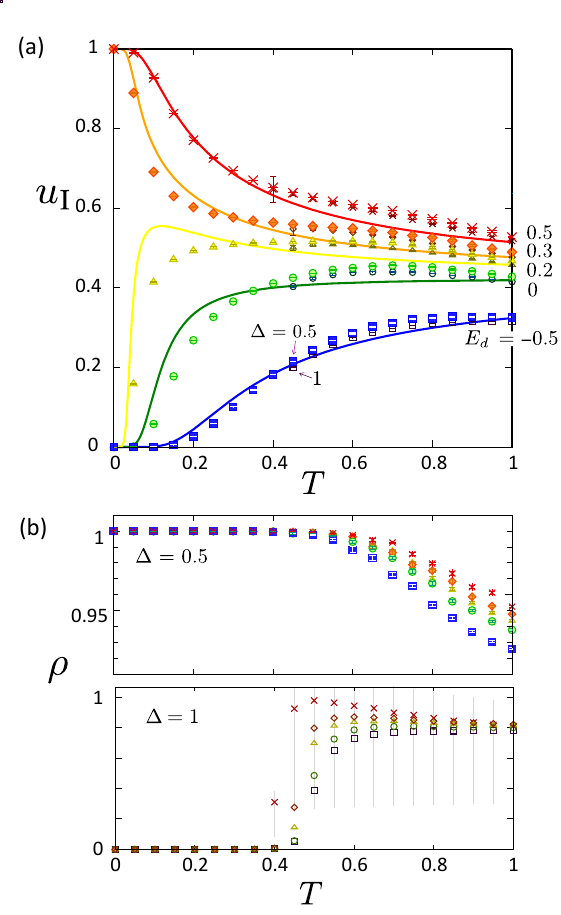}
\caption{
(a) Degree of dimerization scaled by the ionic domain size, $u_{\rm I}$, 
as a function of $T$ for $\Delta=0.5$ and $1$ and various choices of $E_d$. 
The exact solution for the case excluding the N state is given in Appendix~\ref{sec:exactsol}, 
whose results are shown in solid lines. 
(b) Ionicity, $\rho$, for $\Delta=0.5$ (upper) and $1$ (lower panel). 
The error bars on the lower panel for $E_d=0.5$ are shown in gray lines, 
while others are abbreviated for clarity (which are of the same order). 
  }
\label{fig:dimerization}
\end{figure}
\par
To take a closer look at the nature of these phases, we plot $u_{\rm I}$ and $\rho$ 
in Fig.~\ref{fig:dimerization} 
as functions of temperature for several choices of $E_d$ for $\Delta =0.5$ and $1$. 
At zero temperature, there is a discontinuous change between the fully dimerized and nondimerized ionic phases; 
$u_{\rm I}=1$ to $0$ at $E_d>2J=0.25$ and $E_d<2J$, respectively. 
\par
When $E_d \lesssim 2J$, the low temperature phase has $0\le u_{\rm I}\ll 1$. 
Namely, when $\Delta=0.5$, most of the sites are ionic and few isolated dimers are activated thermally, 
which is the case schematically shown in Fig.~\ref{fig:model}(b-iv). 
This excitation does not carry a mobile charge. 
Most of the NIDW phase particularly at small $\Delta$ in Fig.~\ref{fig:phase_diagram}, 
and the state observed in Fig.~\ref{fig:dimerization}(a,b) at $\Delta=1$ and $T\gtrsim 0.5$, 
belongs to this phase. 
\par
When $E_d \gtrsim 2J$ the number of dimerized pairs of sites increases gradually with degreasing $T$. 
In such case, when $\Delta \ll 1$ the system is in ``the ionic phase" with $\rho=1$, but in reality, 
quite many of the sites are dimerized at low temperatures and we expect Fig.~\ref{fig:model}(b-iii) 
with an SS and CS as the primary excitations, both not moving around much as 
we find in Fig.~\ref{fig:trajectory}(d). 
Whereas when $\Delta \gg 1$, we find the neutral phase at low temperature, 
where only a few numbers of dimers are excited as thermal fluctuations in the sea of neutral sites, 
shown in Fig.~\ref{fig:model}(b-i). 
\par
We also plot $u_{\rm I}$ from the exact solution of the model 
that excludes the N state in Appendix~\ref{sec:exactsol}. 
The overall tendency is well reproduced. 
The difference between the two indicates that there is a finite contribution from the N state, 
and also the finite size effect as we take $N=\infty$ in the exact solution. 
\par
\begin{figure}[tbp]
\includegraphics[width=8cm]{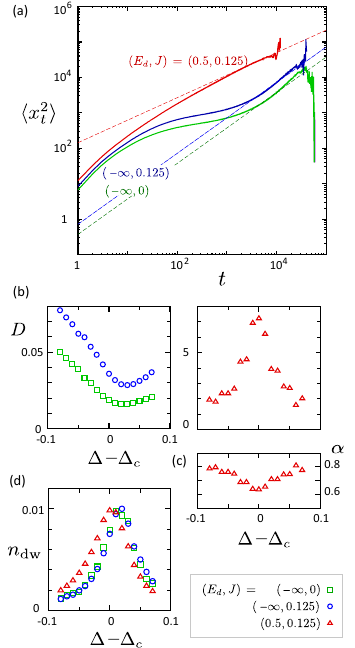}
\caption{
  (a) MSD of NIDWs, $\langle x_t^2\rangle$, as a function of the elapsed time $t$ after their creation for three cases; 
  one with dimerization $E_d=0.5$ and spin-spin interactions $J=0.25$ at $(T,\Delta)=(0.5,1.07)$, 
  the other two with dimerization being absent, $E_d=-\infty$, 
  while with $J=0.25$ at $(T,\Delta)=(0.47,1.17)$, and $J=0$ at $(T,\Delta)=(0.47,1.14)$,  
  where we chose the parameters at the triple points. 
  Broken lines are $\langle x_t^2\rangle=2D x^\alpha$ with $\alpha$ 
  with $\alpha=0.6$ for the subdiffusive case and $\alpha=1$ for the latter two diffusive cases. 
  (b) Values of $D$ obtained by fitting $\langle x_t^2\rangle$, 
  (c) Variation of power $\alpha$ for the subdiffusive case, and 
  (d) Number density $n_{\rm dw}$, as functions of $\Delta-\Delta_c$ with $\Delta_c$ giving $\rho=0.5$. 
}
\label{fig:nidw_dynamics}
\end{figure}
\begin{figure}[tbp]
\includegraphics[width=8cm]{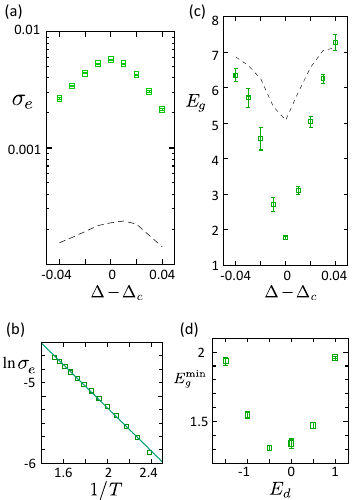}
\caption{
(a) Electrical conductivity $\sigma_e$ calculated under the electric field bias (see \S.{sec:analysis}) 
at around the NI transition point for a set of parameters $E_d=0.5$, $J=0.25$. 
The broken line is the result for ${\cal H}_{\rm d}=0$ as a reference \cite{Sakai2024}. 
(b) $\sigma_e \propto e^{-E_g/T}$ evaluated at the temperature range of $T=0.52 \text{-} 0.58$ 
and (c) the evaluated $E_g$ as a function of $\Delta-\Delta_c$. 
(d) The minimum value, $E_g^{\text{min}}= {\text{min}}_\Delta E_g$ as a function of $E_d$. 
  }
\label{fig:nidw_conductivity}
\end{figure}
%
\subsection{Transport properties of NIDW}\label{subsec:NIDWs_conductivity}
\subsubsection{Diffusion of NIDWs}
We first perform the MC calculation without an applied electric field, 
and by tracking the trajectories of NIDWs, 
calculate the MSD and the number of NIDWs to clarify the nature of their hydrodynamics. 
To understand the effect of dimerization and spin-spin interaction separately, 
we examine three cases; $(E_d,J)=(-\infty,0),(-\infty,0.125),$ and $(0.5,0.125)$, 
where $E_d=-\infty$ means the absence of dimerization 
and $T$ and $\Delta$ are chosen to be at the triple point. 
The time-dependent trajectories of NIDWs for these three cases are shown in Fig.~\ref{fig:trajectory}(a)-\ref{fig:trajectory}(c). 
\par
Figure~\ref{fig:nidw_dynamics} shows the MSD, $\langle x_t^2\rangle$, 
of NIDWs as a function of the elapsed time $t$ after their creation. 
Without dimerization, we find a plateau-like structure, $\langle x_t^2\rangle \sim 100$, between $t=100-1000$ MCSs, 
where the NIDWs are confined within $\sim 10$ sites. 
This phenomenon is observed previously for ${\cal H}_{\rm cl}$ without lattice dimerization effect \cite{Sakai2024}, 
and is attributed to the parallel-spin pairs; 
once the two adjacent electrons on D and A sites each singly occupied have parallel spin orientation, 
they cannot hop to each other due to the Pauli blocking effect, and form a barrier that disturbs the motion of NIDWs. 
The average distribution of the parallel spin pairs is $\sim 10$ sites with a lifetime of $\sim 1000$ MCSs.
In the longer time scale, $\gtrsim 1000$ MCSs, the MSD finally reaches the normal diffusion 
represented by the $t$-linear slope that follows the broken line in the figure. 
When comparing $J=0$ and $0.125$ cases, the plateau of MSD extends to twice and becomes shorter in time, 
because the parallel spin pairs should be destabilized and reduced. 
\par
In contrast, when $E_d=0.5$ and $J=0.125$, the plateau-like structure of MSD disappears, 
and we find $\langle x_t^2\rangle \propto t^\alpha$, 
with $\alpha$ slowly varying from $\alpha \sim 0.7$ toward the subdiffusive $\alpha=0.5$ (dotted line), 
although the present calculation collapses at $t\gtrsim 10^4$ for that parameter. 
The subdiffusive behavior $\langle x_t^2\rangle \propto t^{1/2}$ reminds us of 
a single file diffusion (SFD) \cite{Hahn1996,Wei2000}, for particles that have repulsive interactions. 
In our case, when MSD becomes large, the NIDWs with the same sign of charge start to meet and 
feel the Coulomb repulsive force.  
\par
We fitted $\langle x_t^2\rangle = 2D t^\alpha$ for the three parameters, whose results 
are in Figs.~\ref{fig:nidw_dynamics}(b) and \ref{fig:nidw_dynamics}(c) as functions of $\Delta-\Delta_c$. 
Here, $\Delta_c$ is the parameter that gives $\rho=0.5$, which is phenomenologically regarded as the NI phase boundary. 
For the case without dimerization effect, we set $\alpha=1$ and extract $D$ as a diffusion constant, 
and from these data, can evaluate the conductivity $\sigma_{\rm dw}$ using Einstein relation in Eq.~(\ref{eq:sigmadw}). 
For the subdiffusive case, such a relationship does not hold, whereas the extracted value of $D$ may reflect 
the degree of subdiffusion, that can be compared quantitatively with other results. 
Indeed, the amplitude of $D$ becomes one order of magnitude larger by the dimerization effect 
and shows a peak at $\Delta\sim \Delta_c$ in contrast to the dip in $D$ observed for 
the nondimerized diffusive cases. 
\par 
We also plot the number density $n_{\rm dw}=N_{\rm dw}/N$ of NIDWs in Fig.~\ref{fig:nidw_dynamics}(d), 
which gives the estimate of the carrier density that contributes to the conductivity. 
They all behave similarly, while the peak position shifts to the neutral side when the dimerization is absent, 
as previously observed for the analysis on ${\cal H}_{\rm cl}$ without the lattice degrees of freedom. 
These results show that the nature of conductivity changes both qualitatively and quantitatively 
when the dimerization is introduced, and the major cause would be the change in the types of diffusions. 
\par
\subsubsection{Conductivity and activation gap}
We now calculate the conductivity $\sigma_e$ at around the NI transition point under the electric field bias. 
As shown in Fig.~\ref{fig:nidw_conductivity}(a), $\sigma_e$ takes a maximum at $\Delta=\Delta_c$, 
regardless of the choice of temperatures, $T=0.40\text{-}0.66$, 
and the amplitude of $\sigma_e$ is enhanced by one order of magnitude when ${\cal H}_{\rm d}$ is included. 
By referring to Fig.~\ref{fig:nidw_dynamics}, we find that 
$\langle x_t^2\rangle \propto D$ is indeed by one orders enhanced for the case with $E_d=0.5$ compared to $E_d=-\infty$. 
whereas, carrier number, as mentioned earlier, does not change much. 
Therefore, one can attribute the origin of the suppressed $E_g$ to the enhancement of electron fluctuation, 
or equivalently, the suppression of diffusion. 
Indeed, the fluctuation/mobility of electrons between pairs of sites 
is possibly enhanced before or after the dimerization during MCS with the aid of $E_d$. 
There, the two adjacent electrons need to be antiparallel, reducing the probability of the Pauli blocking effect. 
\par
From the temperature dependence of $\sigma_e$, one can derive the activation gap, $E_g$, 
using the data in the range of $T=0.52 \text{-} 0.58$, e.g. in Fig.~\ref{fig:nidw_conductivity}(b). 
Figure \ref{fig:nidw_conductivity}(c) shows $E_g$ as a function of $\Delta$, 
which takes a minimum at $\Delta=\Delta_c$. 
This value, $E_{g}^{\rm min}=\text{min}_{\Delta} E_g$, is plotted as a function of $E_d$ 
in Fig.~\ref{fig:nidw_conductivity}(d), which takes the minimum at $E_d\sim -0.5-0.0$. 
According to Fig.~\ref{fig:dimerization}, at this parameter range, the degree of dimerization in the ionic domain 
takes the intermediate around $u_{\rm I}\sim 0.25-0.45$ at $T=0.52$, 
i.e. moderately dimerized to less than $50\%$. 
This means that the ionic and dimerized domains coexist as in Fig.~\ref{fig:model}(b-ii), 
and the system is in a highly inhomogeneous phase, and the conductivity should be maximized 
with the aid of such domain structures. 
This result is consistent with the early QMC study on the IEHM including the quantum transfer integral, 
where the electrical conductivity gap is reduced by the electron-lattice coupling at the NI transition point \cite{Nagaosa1986-3}.
%
\subsection{Ferroelectricity under the inter-chain effect}
\begin{figure}[tbp]
\includegraphics[width=8cm]{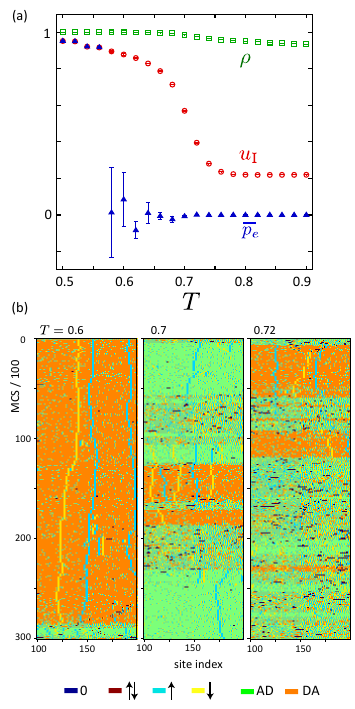}
\caption{Results for ${\cal H}={\cal H}_{\rm cl}+{\cal H}_{\rm d}+{\cal H}_{p}$ 
including the inter-chain dipole-dipole interaction as a mean-field effect. 
We take $\Delta=0.5$, $J=0.125$, $J_e=3$, $\chi=0$ for $N=256$. 
(a) Ionicity, $\rho$, dimerization scaled by the ionic domain size, $u_{\rm I}$, 
and the electric polarization, $p_e$, carried by dimers. 
(b) Snapshots of every 100 MCS for $T=0.72, 0.7, 0.65$. 
The green and orange region mark the AD and DA dimers, and 
blue, red, light blue, and yellow are the neutral and ionic sites. 
  }
\label{fig:ttfba}
\end{figure}
In the experimental pressure-temperature phase diagram of TTF-CA, 
the system undergoes a paraelectric-to-ferroelectric 
transition at the pressure range of $\gtrsim 10$\,kbar in lowering the temperature. 
This transition accompanies the dimerization of basically all pairs of DA molecules. 
The system exhibits a finite electric polarization when the site-centered inversion symmetry is broken, 
namely when having a coherent and static dimerization. 
In the present framework, however, the dimerization is taken into account as a classical 1D degree of freedom, 
which does not exhibit a long-range order at finite temperature. 
In the material systems, however, there is always a finite three-dimensional (3D) effect, in particular for the 
ferroelectric systems, in the form of dipole-dipole interactions. 
\par
We thus take account of the 3D effect by adding an extra term to the Hamiltonian as 
${\cal H}={\cal H}_{\rm cl}+{\cal H}_{\rm d}+{\cal H}_{\rm p}$ with 
\begin{equation}
{\cal H}_{\rm p}=\sum_{l\in\text{dimer}} -J_e  p_l \, \bar p_e , 
\end{equation}
where we set the dipole operator as $p_{l}=+1$ and $-1$ for AD and DA dimers, respectively, 
and $J_e$ is the coupling constant between the local dipole and 
the mean polarization field from the nearby chain, $\bar p_e$. 
In the MC simulation, we treat the polarization field as a uniform mean field averaged over a single chain, 
and assuming that if there are multiple chains that interact with each other, 
their polarization is the same, we adopt this value from the configuration in the previous MCS; 
\begin{equation} 
\bar p_e= \frac{2}{N} \Big({\sum_{j,j+1\in \text{AD}} b_{j,j+1}}-{\sum_{j,j+1\in \text{DA}} b_{j,j+1}}\Big),
\label{eq:pe}
\end{equation}
where AD (DA) represents the bond between AD (DA) site pair, 
and $\bar p_e$ ranges from $-1$ to $1$. 
\par
Figure~\ref{fig:ttfba}(a) shows the ionicity $\rho$, 
the degree of dimerization $u_{\rm I}$ inside the ionic domain, 
and the electric polarization $p_e$ with the input parameter of 
$\Delta=0.5$, $J=0.125$, $E_d=-1$ and $J_e=3$. 
For this choice of parameter, the equilibrium state is almost fully ionic. 
The ferroelectric transition occurs at $T_c\sim 0.56-0.60$, observed as a discontinuity in $\bar p_e$, 
which may indicate the first-order nature. 
As mentioned, this transition is due to the 3D effect introduced by the mean field. 
However, quite importantly, there is a strong fluctuation of $\bar p_e$ at $T\gtrsim T_c$,  
although we do not necessarily expect a criticality typical of a second-order transition. 
One also finds that upon cooling, $u_{\rm I}$ first increases rapidly within $T=0.65-0.7$ and then 
gradually below the temperature, where $\bar p_e$ starts to polarize. 
\par
Figure~\ref{fig:ttfba}(b) shows the snapshots of the states ranging from site 100 to 199 
for every 100 MCS for $T=0.65, 0.7, 0.72$ over $3\times 10^4$ MCS, 
where the green and orange region marks the dimerized AD and DA domains, respectively. 
From these snapshots, we see that there are two free-energy minima, 
which are the states dominated by DA and AD, respectively, and the timescale to 
go back and forth between these minima rapidly increases 
even between $T=0.72$ to $0.7$. Once $u_{\rm I}$ increases up to $\gtrsim 0.6$ 
at $T=0.65$ we could no longer capture the transfer between the minima within the 
timescale plotted. 
The slowing down of domain motions results in the emergent $\bar p_e\ne 0$ at $T\gtrsim T_c$.

\begin{figure}[tbp]
\includegraphics[width=8.5cm]{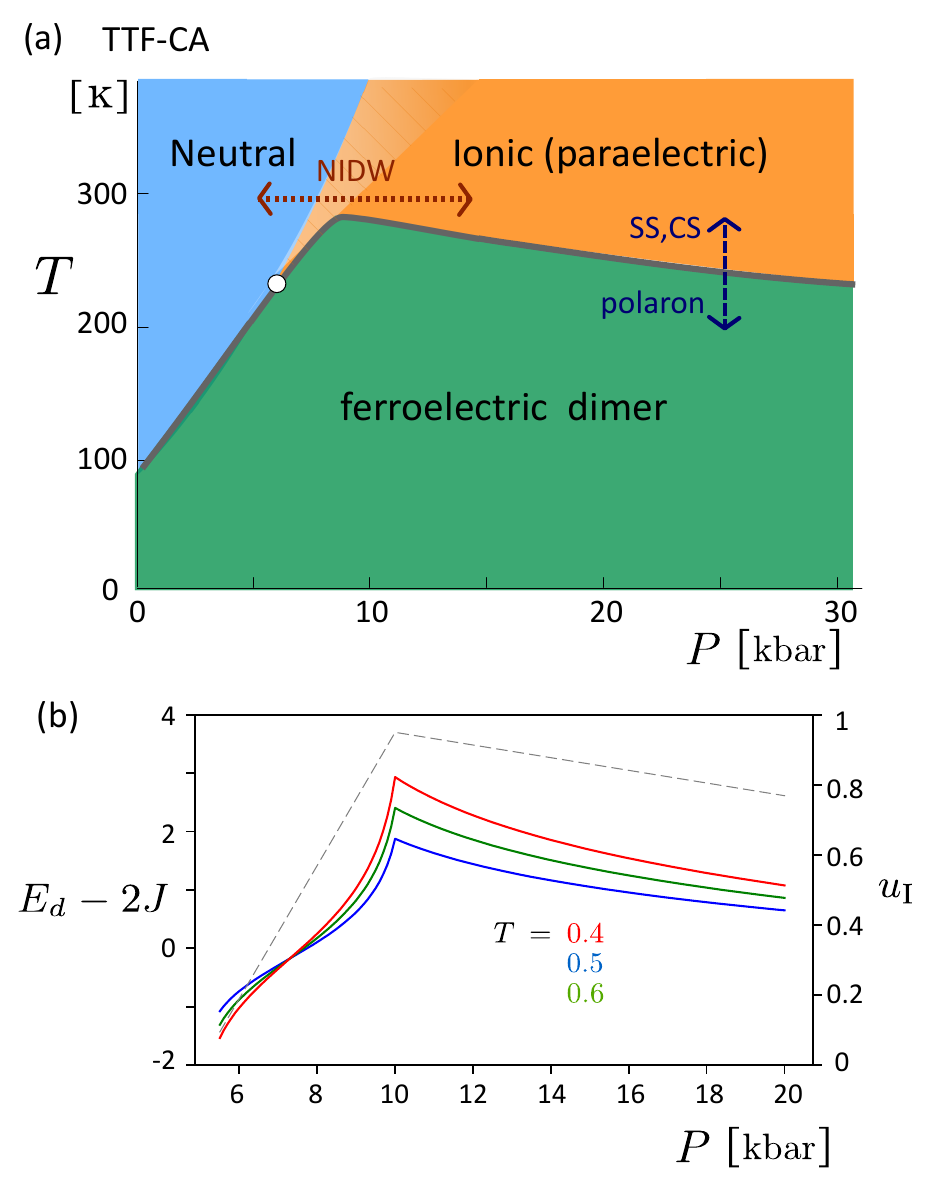}
\caption{
(a) Schematic illustration of the $P$-$T$ phase diagram ofTTF-CA. 
The triple point that connects the N, I, and NIDW phase in Fig.~\ref{fig:phase_diagram} 
is supposed to correspond to the point marked in the figure, 
and the NIDW may represent the paraelectric ionic phase of the phase diagram, 
where the decrease in $\Delta$ may correspond to the application of $P$. 
(b) Dimerization energy gain, $E_d-2J$ (with fixed $J=0.125$), in our model related to 
the pressure effect of the phase diagram by reading off the value of $u_{\rm I}$ 
as the IR peak intensity \cite{Masino2007}. See also Appendix~\ref{sec:exactsol}. 
  }
\label{fig:exp}
\end{figure}
\section{Discussion and conclusion}
\label{sec:discussion}
Let us try to access the experimental $P$-$T$ phase diagram of TTF-CA, 
shown schematically in Fig.~\ref{fig:exp}(a). 
The neutral, ionic (paraelectric), and dimerized ferroelectric phases in the experiment 
may correspond to our N, NIDW, and I phase with dimerization 
in Fig.~\ref{fig:phase_diagram}, 
and since the three phases meet at the triple point, we expect it to locate 
at the position marked in a circle. 
In previous theories, it is shown that a decrease in $\Delta$ drives the 
N-to-I transition, roughly corresponding to the application of pressure in the experiment \cite{Nagaosa1986-1,Nagaosa1986-2}. 
At the crossover region, $P\sim 8.6-12.4$\,kbar \cite{Masino2007}, the NIDWs should be responsible for the transport. 
The electrical conductivity of TTF-CA takes the maximum value at 8.8\,kbar \cite{Takehara2019}. 
Our conductivity $\sigma_e$ shown in Fig.~\ref{fig:nidw_conductivity} is enhanced at 
around the NI transition point $\Delta\sim \Delta_c$, which may correspond to this point, 
and $\sigma_e$ is one order of magnitude enhanced by the introduction 
of the dimerization effect. 
Indeed, the experiments indicate the importance of lattice dimerization 
in this crossover region. 
\par
{\it Pressure dependent dimerization. }
We first focus on the $P=5-15$\, kbar range including NI crossover. 
The pressure dependence of the degree of dimerization of TTF-CA at room temperature was discussed 
based on the infrared (IR) spectroscopy measurement \cite{Masino2007}, 
where the intensity of the normalized IR shift qualitatively represents the lattice distortion amplitude. 
The intensity becomes nonzero at $P\gtrsim 5$\,kbar and increases rapidly up to 
$P\sim 11$\,kbar, and then gradually decreases. 
\par
In the early theoretical work based on exact diagonalization of the strong coupling model 
an abrupt increase of the degree of lattice distortion at the N-to-I transition was observed, 
parameterized by $V=-2\Delta +$ constant \cite{Soos2007} or by the ionicity $\rho$ \cite{Painelli1988}. 
After entering the I phase, the lattice distortion takes a maximum and decreases gradually. 
Because $\Delta$ is read off as negative pressure in the experiment, this result seems to successfully 
explain the experimental findings. 
The lattice distortion amplitude may take a maximum value at the point where the 
charge fluctuation between N and I sites is maximized. 
Their dimerization energy gain, read off as our $E_d$, 
indeed coherently follows the increase-to-decrease behavior in lowering $\Delta$. 
The early QMC calculation is consistent with this result \cite{Nagaosa1986-3,Otsuka2012}. 
\par
However, in our case, $E_d$, and $\Delta$ are taken as independent parameters. 
Within this framework, $u_{\rm I}$ is almost independent of $\Delta$ (see Fig.~\ref{fig:dimerization}(a)). 
Particularly, when confined to either the ionic or dimerized sites, 
$u_{\rm I}$ depends only on $E_d-2J$ and $T$ (see Appendix \ref{sec:exactsol}), 
which are responsible for the energy gain and entropic gain, respectively. 
Remind that the microscopic origin of $E_d$ is the energy gain due to singlet formation 
and the lattice elastic energy loss added on top. 
\par
Combining our results with the experimental observation,  
we extract $E_d$ as a function of $P$ in Fig.~\ref{fig:exp}(b); 
We set $T=0.4,0.5,0.6$ and $\Delta=1$ that may correspond to the crossover range in Fig.~\ref{fig:exp}(a). 
The $P$-dependence of normalized IR peak intensity increases almost linearly from 0 to 1 as $\sim (P/5-1)$ 
at $P=5-10$\,kbar and then becomes flat until 15 kbar and then slightly decreases. 
We adopt the maximum degree of dimerization in the ionic phase independently measured, 
and set $u_{\rm I}=0.95(P/5-1)$ for $P\le 10$\,kbar and otherwise slowly decreasing at $P\ge 10$\,kbar, 
as shown in the broken line. 
We compare this value with the exact solution obtained in the Appendix~\ref{sec:exactsol}
for the ionic phase obtained by excluding the N site configuration, 
and obtain the $P$ dependence of $E_d-2J$ with $J=0.125$ being fixed. 
One finds indeed that $E_d-2J$ increases rapidly, taking a maximum, and then starts to decrease slowly. 
\par
{\it Charge soliton excitation in the paraelectric ionic phase. } 
At the pressure range of $P\gtrsim 15$\,kbar in the paraelectric ionic phase, 
experiments suggest that the excitation of TTF-CA is dominated not by NIDWs but by the spin(SS) and charge(CS) solitons \cite{Takehara2023}. 
The activation energy gap of the electrical conductivity attributed to such excitation is evaluated as 
$0.06- 0.08 \rm eV$ \cite{Takehara2023}.
In our calculation, it corresponds to the I phase with $\rho\sim 1$  
as depicted schematically in Fig.~\ref{fig:model}(b-iii). 
From Fig.~\ref{fig:exp}(c), $E_d-2J$ ranges from around 0.5 down to $0$ with increasing $P(\gtrsim 15$\,kbar), 
which reduces $u_{\rm I}$ from 0.8 to 0.5, namely the dimerization is weakened. 
\par
This explains the decrease of the ferroelectric transition temperature $T_c$ with increasing $P$. 
As we also found in Fig.~\ref{fig:nidw_conductivity}(d) that 
within this range, the minimum activation gap $E_g^{\text{min}}$ of the conductivity $\sigma_e$ 
is reduced with decreasing $E_d$. 
Because $E_g^{\text{min}}$ gives the excitation energy of CS in the high-pressure range, 
the conductivity becomes larger with pressure, coherently with the weakening of dimerization. 
Experimentally, the electrical conductivity is rather insensitive to pressure up to $P\sim 82.5 \rm kbar$ \cite{Takehara2015}, 
which is unusual as an organic semiconductor, 
while the activation gap gradually decreases from $0.08 \rm eV$ to $0.06 \rm eV$ 
comparable to those of the crossover region \cite{Takehara2023}.
It was then argued that the increase of transfer integral $t_r$ due to pressure is responsible \cite{Takehara2023}. 
However, since the transport here is presumably not carried by the electrons itself but rather by the CS,  
which is the collective motion of a few electrons with the aid of the fluctuation of dimers, 
the degree of dimerization energy gain is the key factor to understand the overall feature of the phase diagram. 
\par 
{\it TTF-BA. }
We finally refer to the other family member of the material, TTF-BA. 
It is an ionic crystal hosting only paraelectric and ferroelectric phases. 
The experimentally evaluated ionicity is $\rho\sim 0.95$ \cite{Girlando1985,Garcia2005},  
which is much larger than the saturated ionicity of $\rho\sim 0.7$ of TTF-CA. 
Unlike TTF-CA, TTF-BA exhibits little to no dimerization at room temperature. 
However, with lowering the temperature at ambient pressure, 
dimerization gradually develops locally as a precursor to the phase transition, 
as demonstrated by NMR and NQR experiments \cite{Sunami2020,Sunami2021-2}. 
With these distinct differences, 
although TTF-BA was first expected to represent the high-pressure state of TTF-CA, 
they are not necessarily connected to each other. 
\par
In our results, Fig.~\ref{fig:ttfba}, including the inter-chain interaction, 
there is a certain amount of fluctuation between the DA and AD domains carrying different $p_e$ in our simulation 
at temperatures slightly above the ferroelectric transition. 
We consider that such an effect essentially captures the local dimerization observed as a 
precursor to the ferroelectric transition in TTF-BA \cite{Sunami2020}. 
\par
When lowering the temperature and turning on the transfer integral $t_r$ to go back to the IEHM, 
the strong thermal fluctuation of $\bar p_e$ 
is converted to the mixture of N and I configuration. 
We show in Appendix~\ref{sec:twosite} 
the solution of the two-site IEHM model including $t_r$ only inside the dimer, 
and taking account of the intersite $V$ as mean-field. 
There is a first-order NI transition when $V$ is large, consistent with the previous theories \cite{Nagaosa1986-2}. 
The parameter controlling the ionicity is $(U-\Delta-1.5V)/t_r$, which changes much with $t_r$. 
In this context, one can consider the dimerized ferroelectricity as 
being the electron-type and the lattice-displacement type, 
which takes the charge-neutral state before turning on the ferroelectricity 
as the N and I phases, respectively. 
The mixture of N and I configurations increases gradually with $t_r$, 
so that the wave function of the two cases is indeed subtle, 
and the lattice displacement is the only degree of freedom that breaks the symmetry 
and marks the ferroelectric transition. 
\par
{\it Conclusion. }
To summarize, we studied the classical model relevant to the organic compounds TTF-CA and TTF-BA, 
which exhibit neutral-ionic transitions. 
Our model corresponds to the strong-coupling limit ($t_r \rightarrow 0$) 
of the ionic extended Hubbard model with on-site ($U$) and nearest-neighbor ($V$) Coulomb interactions, 
defined on a one-dimensional lattice composed of donor (D) and acceptor (A) sites 
with an energy difference $\Delta$, incorporating local lattice dimerization effects. 
Previously, we showed that in the absence of lattice distortion, 
a neutral-ionic domain wall (NIDW) phase emerges near the neutral-to-ionic crossover, 
enhancing the electrical conductivity $\sigma_e$ through the diffusive motion of domain walls \cite{Sakai2024}. 
In this study, we extended the model to include stochastic lattice dimerization; 
the ionic site carrying one electron with either up or down spin 
interacts with its neighbors antiferromagnetically by $J$, 
while the two neighboring ionic sites can form dimers with an associated energy gain $E_d$. 
Our Monte Carlo simulations reveal that the degree of dimerization is controlled by the 
thermal fluctuation and the competing energy gain from $E_d$ and $J$, 
and the dimerization becomes prominent when $E_d > 2J$. 
The finite population of ionic and dimerized sites yields 
locally fluctuating spin and dimer configurations that promote subdiffusive electronic dynamics 
and a significant enhancement of conductivity at the crossover. 
In the highly ionized regime, dimerization fluctuations also increase 
the ferroelectric transition temperature via an entropic stabilization mechanism. 
By interpreting external pressure as modulating $E_d$, we qualitatively account for 
the experimental phase diagrams of TTF-CA and TTF-BA.

\appendix
\section{Exact solution of ${\cal H}_{\rm d}$ at $\rho=1$} 
\label{sec:exactsol}
\begin{figure}[tbp]
\includegraphics[width=8cm]{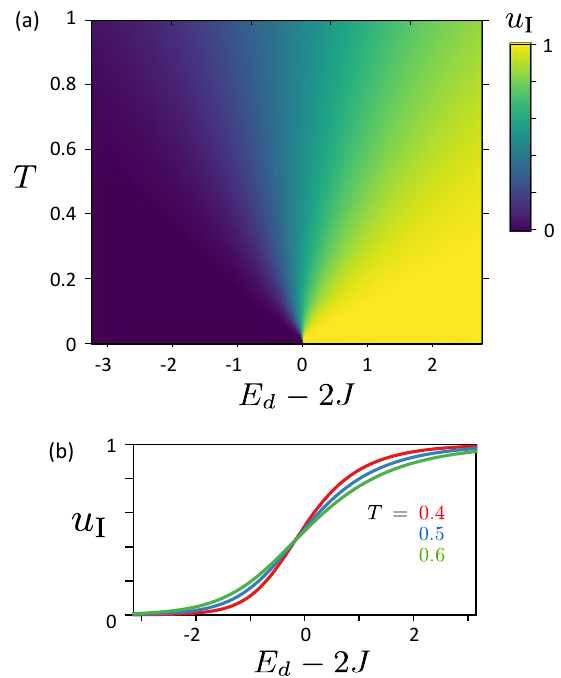}
\caption{
The exact solution obtained by the transfer matrix method for our model 
without neutral configuration. 
$u_{\rm I}$ depends on $E_d-2J$ and $T$, plotted for several choices of f...
  }
\label{fig:exact}
\end{figure}

We consider the case where all the sites either belong to the ionic state or join the dimerization. 
In this simplified model, given formally as ${\cal H}_{\rm d}$ in Eq.~(\ref{eq:hamd}), 
we have one electron per site, and the system reduces to an antiferromagnetic 
Ising spins $\sigma_i=\pm 1$ coupled to the dimerization degrees of freedom. 
One can perform a transfer matrix method to obtain the partition function and 
the thermodynamic properties at $N\rightarrow\infty$. 
For this purpose, we redefine a parameter $\tau_i$ on-site $i$ as 
\begin{equation}
\tau_i=\left\{
\begin{array}{rl}
 1   &:\quad \text{sites } i \text{ and } i{+}1 \text{ form a dimer} \\
-1   &:\quad \text{sites } i{-}1 \text{ and } i \text{ form a dimer} \\
 0   &:\quad \text{no dimerization of a spin at site } i
\end{array}
\right.
\end{equation}
Then, we combine the two variables on-site $i$, which spans the local Hilbert space as
\begin{align}
 (\sigma_i,b_i) 
&= (+1,0), (-1,0), (+1,1), (-1,1), \notag \\
&\quad (+1,-1), (-1,-1), 
\end{align}
and with this basis, the $6\times 6$ transfer matrix $T$ is given as 
\begin{equation}
  T =
  \begin{pmatrix}
    e^{-\beta J} & e^{\beta J} & 1 & 1 & 0 & 0 \\
    e^{\beta J} & e^{-\beta J} & 1 & 1 & 0 & 0 \\
    0 & 0 & 0 & 0 & 0 & e^{\beta E_d} \\
    0 & 0 & 0 & 0 & e^{\beta E_d} & 0 \\
    1 & 1 & 1 & 1 & 0 & 0 \\
    1 & 1 & 1 & 1 & 0 & 0
  \end{pmatrix},
  \label{eq:transfer_matrix}
\end{equation}
where $\beta$ is the inverse temperature. 
The secular equation of $T$ is reduced to 
\begin{align}
&  \lambda^2 \{\lambda + 2\sinh(\beta J)\} \times \notag\\ 
&  \{\lambda^3 - 2\cosh(\beta J)\lambda^2- 2e^{\beta E_d}\lambda +4e^{\beta E_d}(\cosh(\beta J)-1)\} = 0.
  \label{eq:secular}
\end{align}
It can be straightforwardly shown that for all $E_d$ and $J>0$, 
the root of Eq.~(\ref{eq:secular}) having the largest absolute value 
is simultaneously the root of the cubic equation in the last parenthesis with the maximum positive value. 
Using this root, $\lambda_{\text{max}}(E_d,J,\beta)$, the partition function is given as 
$Z = \lambda_{\text{max}}^N$, and the degree of the dimerization $u_{\rm I}$ is derived as,
\begin{equation}
  u_{\rm I}(E_d,J)
  = \frac{2}{\beta}\frac{\partial}{\partial E_d}
    \ln \lambda_{\text{max}},
  \label{eq:dimerization_derivation}
\end{equation}
We fixed $J=0.125$ and numerically calculated $\lambda_{\text{max}}$ 
and its derivative in terms of $E_d$. 
The density plot of $u_{\rm I}$ on the plane of $E_d-2J$ and $T$ is shown in Fig.~\ref{fig:exact}(a). 
The $E_d$ dependence of $u_{\rm I}$ for $T=0.4,0.5,0.6$ are 
shown in Fig.~\ref{fig:exact}(b), which is used to derive the pressure dependence of $E_d$ in Fig.~\ref{fig:exp}(b) 
in comparison with the experimental observations. 


\begin{figure}[tbp]
\includegraphics[width=8cm]{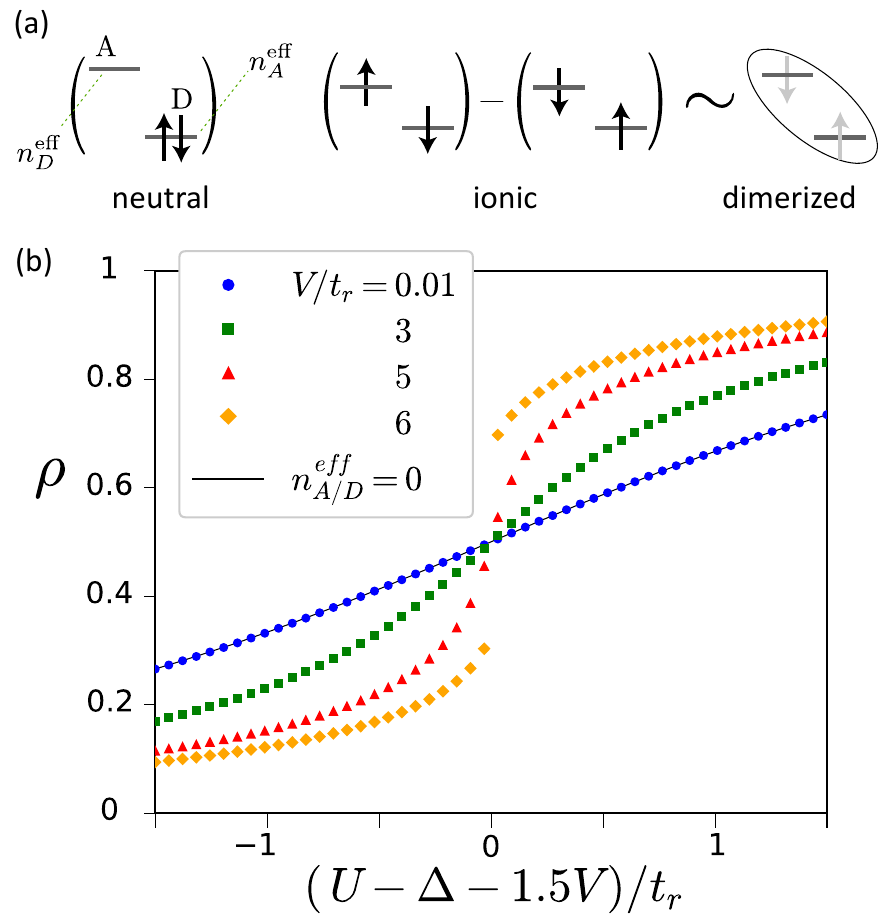}
\caption{
Schematic illustration of the two-site model with neutral and ionic configuration, 
and the self-consistent solution including the mean-field from the other neighboring sites. 
  }
\label{fig:app}
\end{figure}
\section{Quantum ionic extended Hubbard model for the dimerized two sites}
\label{sec:twosite}
To see how the neutral and ionic states may change with $\Delta$ in the presence of quantum hopping, 
we consider the modified IEHM defined on two sites, 
\begin{align}
{\cal H}_{\rm IEHM}&= \sum_\sigma (-t_r c_{A\sigma}^\dagger c_{D\sigma} +\text{h.c.}) + \frac{\Delta}{2}(n_A-n_D)
\notag\\
&+ \sum_{j=A,D} U n_{j\uparrow}n_{j\downarrow} 
+ V (n_An_D + \frac{1}{2}(n_An_D^{\rm eff} + n_A^{\rm eff} n_D)), 
\end{align} 
where the operators have indices D and A. 
We consider the case of half-filling and exclude the double occupancy of A sites for simplicity. 
Here, $n_{A/D}^{\rm eff}$ denote the mean-field effect from the sites from outside the dimer, 
which is solved self-consistently. 
When setting $n_{A/D}^{\rm eff}=0$, the ground state of an isolated dimer is given as 
\begin{align}
|\psi_{\rm gs}\rangle
&=\frac{1}{\sqrt{1+e^{-2\theta}}}
\big(e^{-\theta}|0,\uparrow\downarrow\rangle + \frac{1}{\sqrt{2}}
(|\uparrow,\downarrow\rangle - |\downarrow,\uparrow\rangle\,) \big),\notag\\
& \sinh\theta=\frac{1}{2\sqrt{2}}(u-v),
\end{align}
where $u=(U-\Delta)/t_r$ and $v=V/t_r$. 
The neutral $|0,\uparrow\downarrow\rangle$ and 
ionic $|\uparrow,\downarrow\rangle - |\downarrow,\uparrow\rangle$ states
mix with increasing $V$ or $\Delta$, 
and the ionicity is evaluated as $\rho(\theta)=(1+e^{-2\theta})^{-1}$. 
Figure~\ref{fig:app}(a) shows schematically the neutral and ionic states 
that constitute $|\psi_{\rm gs}\rangle$. 
The superposition of the two ionic states become nonmagnetic and is 
equivalent to the quantum dimerized state with $\rho=1$. 
By introducing the mean-field term, the self-consistent solution is obtained numerically, 
as shown in Fig.~\ref{fig:app}(b). 
where we find the first order transition at large enough $V$. 
The transition takes place at $U-\Delta=1.5V$ which is slightly smaller than 
the fully classical case $U-\Delta=2V$. 
\par
Importantly, we put only $t_r=1$ between a single pair of A and D, 
and $(U-\Delta-1.5V)$ is scaled  by $t_r$, meaning that setting $t_r \rightarrow 0$ 
will make $\rho$ more rapidly increase. 
When $U$ is sufficiently large and $\rho\sim 1$ the wave function describes 
the dimerized ionic state. 
The only difference from the neutral state is whether the center of charge 
is located on D or at the center of the bond. 
\par
In this context, the electronic ferroelectricity could be picturized as 
$\rho\lesssim 1$ where a finite amount of neutral configuration participates. 
The degree of such electronic component may depend much on $t_r$ against other parameters; 
this may explain the difference between the ferroelectric nature of 
TTF-CA and TTF-BA.

\section*{Acknowledgments}
We thank Prof. Kazushi Kanoda and Dr. Keishi Sunami for the discussions. 
This work was supported by JST SPRING, Grant Number JPMJSP2108, 
"The Natural Laws of Extreme Universe"(No. JP21H05191) KAKENHI for Transformative Areas from JSPS of Japan,
 and JSPS KAKENHI Grants No. JP21K03440. 
\bibliography{phd_thesis}
\end{document}